\documentclass[aps,preprint,floatfix,%
showpacs,showkeys,amssymb]{revtex4}

\usepackage{graphicx}

\begin{document}

\date{\today} 

\def\ba{\begin{array}}
\def\ea{\end{array}}
\def\be{\begin{equation}\begin{array}{l}}
\def\ee{\end{array}\end{equation}}
\def\bea{\begin{equation}\begin{array}{l}}
\def\eea{\end{array}\end{equation}}
\def\f#1#2{\frac{\displaystyle #1}{\displaystyle #2}}
\def\om{\omega}
\def\omm{\omega^a_b}
\def\we{\wedge}
\def\de{\delta}
\def\De{\Delta}
\def\va{\varepsilon}
\def\omb{\bar{\omega}}
\def\la{\lambda}
\def\vv{\f{V}{\la^d}}
\def\si{\sigma}
\def\t{T_+}
\def\v{v_{cl}}
\def\m{m_{cl}}
\def\n{N_{cl}}
\def\bi{\bibitem}
\def\c{\cite}
\def\sa{\sigma_{\alpha}}
\def\ua{\uparrow}
\def\da{\downarrow}
\def\mua{\mu_{\alpha}}
\def\ga{\gamma_{\alpha}}
\def\g{\gamma}
\def\G{\Gamma}
\def\ora{\overrightarrow}
\def\pa{\partial}
\def\ov{\ora{v}}
\def\al{\alpha}
\def\bt{\beta}
\def\R{R_{eff}}
\def\th{\theta}
\def\na{\nabla}

\def\muu{\f{\mu}{ed}}
\def\E{\f{edE(\tau)}{\om}}
\def\t{\tau}

\baselineskip 6.2mm
\parskip 1.5mm

\title{Vortices and Dissipation in a Bilayer Thin Film Superconductor}

\author{Wei Zhang and H. A. Fertig}

\affiliation{Department of Physics, Indiana University, Bloomington, IN 47405}

\begin{abstract}
Vortex dynamics in a bilayer thin film superconductor are
studied through a Josephson-coupled double layer XY model.
A renormalization group analysis shows that there are
three possible states associated with the {\it relative} phase
of the layers:
a free vortex phase, a logarithmically confined vortex-antivortex
pair phase, and a linearly confined phase.  The phases
may be distinguished by measuring the resistance to counterflow current.
For a geometry in which current is injected
and removed from the two layers at the same edge by an ideal 
(dissipationless) lead,
we argue that the three phases yield distinct behaviors: metallic
conductivity in the free vortex phase,
a power law I-V 
in the logarithmically confined phase, and true dissipationless superconductivity
in the linearly confined phase. Numerical simulations of a
resistively shunted Josephson junction model reveal size dependences for
the resistance of this system that support these expectations.
\end{abstract}


\maketitle

\section{Introduction}
Topological defects play an important role in many condensed matter systems.
A paradigm of this are vortices in systems whose energetics may be
described by a single angular variable $\theta({\bf r})$ that is a function of 
position ${\bf r}$,
the simplest example being an $XY$ magnet (also known as the
planar rotor model \cite{kadanoff}.)  The state of the vortices
in two-dimensional systems determines important physical properties.
For example, at high temperatures free vortices in 
superfluid films and thin film superconductors lead to
dissipation.  At low temperatures, the defects are 
bound into vortex-antivortex (VAV) pairs, yielding
a state with power-law (quasi-critical) behavior in the correlation
function $\langle e^{i\theta({\bf r})}e^{-i\theta({0})} \rangle$.
While neither the bound nor the unbound vortex phase
supports long-range order \c{mw}, 
there is a well-known thermodynamic phase transition -- the Kosterlitz-Thouless
transition \c{kt} -- in which the vortices unbind at a critical
temperature $T_{KT}$.

The situation is dramatically changed when 
one introduces an explicit symmetry-breaking field
that aligns the angles (such as a magnetic field in the
$XY$ model).  One of us \c{herb1,herb2,herb3} investigated
this situation recently and found that the vortices have
{\it three} possible phases: a
free vortex phase, a logarithmically bound VAV
pair phase, and a linearly confined phase.
Can these phases be distinguished in an
experimental system described by this model?
This is the issue we address in
this paper.

The problem of the resistive
transition in a single layer superconductor was studied in Ref.
\onlinecite{hn}.
It was found that for $T>T_{KT}$, a current produces a force
on a free
vortex, whose subsequent motion then induces 
a voltage along the direction of current.
This produces
a linear $I-V$ curve. For $T<T_{KT}$, bound VAV pairs have no
{\it net} force on them due to a current and so create no net
voltage drop.  A voltage is induced by
unbound vortices, but these are only induced by the current,
leading to a non-linear
(power law) $I-V$ curve at low currents.  In this work, we study the analog of this
for a system with explicit symmetry-breaking: the voltage
response of
a Josephson-coupled 
bilayer thin film superconductor with respect to an injected counterflow current.
As discussed below, the Josephson coupling acts as an explicit
symmetry-breaking for the {\it relative} phase of the order
parameters for the two layers, $\theta_1({\bf r})-\theta_2({\bf r}).$
Isolated vortices in the relative phase respond to currents in {\it opposite}
directions in the two layers, and their motion induces an interlayer
voltage.  Thus, to probe vortices in the relative phase, 
current must run in opposite directions in the two layers;
i.e., we must have a counterflow current.
A geometry in which this is induced results when current
is injected in one layer and removed from the other at the {\it same}
edge of the sample, which to our knowledge was first studied by Ferrell and Prange \c{cur},
in the absence of vortex excitations.  In this situation, the current
penetrates the system up to a characteristic length scale -- the 
Josephson length -- set by
the tunneling matrix element and the superfluid stiffness of the layers.
Unlike the case for current in an individual layer, the counterflow current can decay 
because an injected charge carrier in the top layer can tunnel
to the bottom layer, and exit the system on the same edge of the 
bilayer from which it entered.

As we show below,
when vortices are introduced, their
dynamics generates different behavior for the three phases:
metallic conductivity in the free vortex phase,
a power law $I-V$ in the logarithmically confined phase,
and true dissipationless superconductivity (zero resistance at
finite current) in the linearly confined phase.
This last phase is qualitatively different from anything that
occurs for the voltage response of a single superconducting layer.
As we demonstrate via simulation below,
by measuring the voltage difference between the two layers at a single edge, we may 
distinguish the three phases. 

The organization of this paper is as follows.
In Section 2, we introduce our model and map it to a Josephson coupled
bilayer XY model. In Section 3, the phase diagram is studied by
duality transformation and a renormalization group (RG) analysis. Section 4
discusses the $I-V$ curves and finite system effects. The numerical
simulations are presented in Section 5, and we conclude with a summary.

\section{Theoretical model}
We consider a bilayer thin film superconductor, which can be described
by a form of the Lawrence-Doniach (LD) model \c{ldm}. The LD free
energy may be written as
\be
{\cal F}=
\int d^2 r \sum_{n=1,2} 
\Bigl\lbrace
A|\psi_n|^2 +\f{B}{2}|\psi_n|^4+
2 \al|\na\psi_n|^2+h\cos(\th_1-\th_2)
\Bigr\rbrace,
\ee
where $\psi_1=|\psi_1|e^{i\th_1},~\psi_2=|\psi_2|e^{i\th_2}$ are the order
parameters for the two layers with coordinates ${\bf r}=(x,y)$, and the last term
is the Josephson coupling between the two layers. Neglecting
fluctuations of $|\psi_n|$, we obtain a free-energy functional with only the phases
involved,
\bea
{\cal F}=\int d^2 r 
\Bigl\lbrace
2\al (\na\th_1)^2+2 \al(\na\th_2)^2+h\cos(\th_1-\th_2) 
\Bigr\rbrace\\
=
\int d^2 r 
\Bigl\lbrace
\al [\na(\th_1+\th_2)]^2+\al[\na(\th_1-\th_2)]^2+h cos(\th_1-\th_2)
\Bigr\rbrace.
\eea
The free-energy ${\cal  F}$ does not support vortex excitations, because
in neglecting variations in the amplitudes of $|\psi_i|$ we did not allow
for the zeros that are necessarily contained in their cores.  Two 
reintroduce these, we replace ${\cal  F}$ with a free energy functional
$F$ whose degrees of freedom are defined on a (square) lattice.
By making the replacement
$1-\f{1}{2}\th^2 \rightarrow \cos(\th)$, we allow for configurations
in which plaquettes of the square lattice may contain a non-vanishing
vorticity.
The resulting free-energy, which has a form similar to what expects for
a bilayer Josephson junction array, is
\be
F=\al \sum_{<r,r'>} \cos \bigl[ \th_1(r)-\th_1(r') \bigr]+
\al \sum_{<r,r'>} \cos \bigl[\th_2(r)-\th_2(r') \bigr]
- h \sum_r \cos \bigl[ \th_1(r)-\th_2(r) \bigr] ,
\ee
where  $<r,r'>$ refers to nearest neighbor sites.  
Models of this form have been studied previously in \c{dxy}.

The analysis of $F$ is complicated by the fact that, in computing the partition
function $Z=\int {\cal D} \theta e^{-F}$, there are cosines appearing in the
exponential.  As is well-known \cite{kadanoff,jose}, considerable
progress can be made if we replace the Josephson coupling form
of the partition function with a Villain model \c{villain}.  This essentially
involves replacing $e^{J\cos \theta}$ with 
$\sum_{m=-\infty}^{\infty}
e^{-J(\theta -2\pi m)^2/2}
$
wherever it appears in the partition function.  The replacement works because
the effective weighting has the same periodicity as the original Josephson
coupling form, so that we should retain the same possible phases for the
system \c{kadanoff}.  We obtain an important simplification by replacing
a cosine form with a Gaussian weighting because it ultimately allows one to
integrate out the angular degrees of freedom \c{jose}.  The cost, however,
is the introduction of effective integer degrees of freedom, which with
some work can be understood in terms of the vortex excitations of the
system \cite{jose,herb1}.  Following a standard trick for rewriting the
partition function as a sum over dual degrees of freedom \cite{jose,herb1},
we find the partition function for the Villain form of the free energy may
be written as
$Z=\sum_{{\bf S},T}e^{-F_{VM}[{\bf S},T]}\Pi_r\de(\na \cdot {\bf S}_1(r)-T(r))
\de(\na \cdot {\bf S}_2 (r)+T(r))$, with
\be
F_{VM}[{\bf S}]=\f{1}{2\al}\sum_r [{\bf S}_1^2(r)+{\bf S}_2^2(r)]+\f{1}{2h}\sum_r T^2(r),
\ee
where ${\bf S}_{1,2}$ are integer fields lying on the bonds of the
lattice in each layer, and $T$ is an integer field defined on the lattice sites.

\section{vortex phase diagram}
A useful representation \c{herb1,herb2}
of the bond degrees of freedom may be
written as
$S_{1x}=\pa_y m,~S_{1y}=-\pa_x m-A,~ S_{2x}=\pa_y n,~S_{1y}=-\pa_x n+A
$, 
where $n$, $m$, and $A$ are all integer fields.
In terms of these the effective free energy $F_D$ for the
partition function $Z=\sum_{m,n,A}e^{-F_D}$ is
\be
F_{D}=\f{1}{2\al}\sum _r|\na m(r)+A(r) \hat{x}|^2+\f{1}{2\al}\sum_r
|\na n(r)-A(r)\hat{x}|^2+\f{1}{2h}\sum_r(\f{\pa A}{\pa y})^2 .
\label{fd}
\ee
As has been discussed elsewhere, the single layer version of
this  -- in which the middle term of \ref{fd} is essentially absent --
may be understood as a model of an interface, with domain
walls and screw dislocations \c{herb1}.  In that situation
the $m$ field allow us to
represent configurations with closed domain walls, and the $A$ field introduces
open domain wall configurations. 
For $F_D$, we have closed domain wall configurations separately introduced
in each layer by the $m,~n$ fields, and the $A$ field, rather than completely
eliminating sections of closed domain walls as in the single layer case,
here allows it to shift between the two layers.  Because of their close
analogy, we will call these ``kinks" in the domain walls dislocations
in the remainder of this paper.  Physically, the domain walls may be
interpreted as worldline trajectories for Cooper pairs,
and the dislocations represent interlayer tunneling events.

One advantage of this representation
is that the coupling constants are the inverse of those in the original functional.
This strong-weak duality makes it convenient to study the physics
with strong inter-layer coupling.
We can obtain further insight by looking at the problem in the vortex representation.
To do this, we apply the Poisson resummation rule \c{jose} on the integer
fields $m,~n$, and arrive another representation of the partition function
in terms of integer fields $M$, $N$ and $A$.
The energy functional associated with these degrees of freedom is
\bea
F_V=-\sum_q \f{2\pi^2K}{S} \f{|M+N|^2}{|{\bf Q}|^2}-\sum_q \f{2\pi^2 K}{S} \f{|H|^2}{|{\bf Q}|^2}-
 \sum_q (\f{1}{2KS|{\bf Q}|^2}+\f{1}{2h}){|Q_y|^2} |A({\bf q})|^2 \\
-\f{2\pi i}{S}\sum_q \f{(Q_x)}{|{\bf Q}|^2} A({\bf q})H^*({\bf q}),
\label{fv}
\eea
where $S$ is the number of lattice sites, $M$ and $N$ are the vortex numbers for the
two layers, $H=M-N$, $K=\al/2$, ${\bf Q}=(Q_x,Q_y)$, and $Q_x=1-e^{-iq_x a}, Q_y=1-e^{-iq_y a}$,
with $a$ the lattice constant.  Physically, we understand the $M+N$ integer field as the vortices
of the symmetric combination of the original layer phases, whereas $H$ represents vortices
for the antisymmetric combination.  
Comparing the terms involving $H$ with
Eq. 12 in Ref. \onlinecite{herb2}, we see that the energetics of anti-symmetric
vortices are identical to those of the single layer $XY$ model with external
magnetic field.  It is thus not surprising that the phases of this
system are analogous to those found in the latter problem.

It is useful to note that in the representation of Eq. \ref{fv},
the partition function shows a near duality.  This can be seen more
clearly if one defines $J({\bf q})=Q_y A({\bf q})$.  In real space
$J$ is also an integer field, and if one eliminates $A$ in favor
of $J$ in Eq. \ref{fv}, then a near symmetry is apparent under
interchange of $J$ and $H$ and $K \rightarrow 1/4\pi K$ in $F_V$. 
To fully exploit this symmetry, we add a term of the form $E_c \sum_q |H({\bf q})|^2$,
after which the duality ($K \rightarrow 1/4\pi K$, $E_c \leftrightarrow 1/h$)
becomes exact \cite{jose}.  The integer fields 
$H$ and $J$ may be respectively interpreted as the vortex field and
a dislocation field, with the added $E_c$ term representing
a core energy for vortices.  The duality in this representation shows that
if we can determine what happens to the dislocations in one part
of the phase diagram, we will know what happens to the vortices in another.

To proceed with finding the phases of this system, we need to perform
an RG analysis.  To do this, we wish to work with continuous rather 
than integer degrees of freedom.  Following the standard reasoning
that the phases depend on the symmetries of the Hamiltonian and
not the precise form of the degrees of freedom \c{cardy}, we 
replace the partition function with one of the form
$Z_{eff}=\int{\cal D} \theta \int{\cal D} \phi \int{\cal D} a e^{-H_{eff}}$,
with

\bea
H_{eff}=\int d^2 r \f{1}{2\al}|\na \phi_1+a\hat{x}|^2+\f{1}{2\al}
|\na\phi_2-a \hat{x}|^2+\f{1}{2h}(\f{\pa a}{\pa y})^2 \\
(-y_{\phi_1} \cos2\pi \phi_1+y_{\phi_1})+(-y_{\phi_2} \cos2\pi \phi_2+y_{\phi_2})
+(-y_{a} \cos2\pi a+y_{a} ).
\eea
Here the last three terms are added to favor
configurations in which $\phi_{1,2}$ and $a$ are integers.
The parameters $y_{\phi_{1,2}}$ have the physical
interpretation of fugacities for vortices in the individual layers, and 
for small values they
favor configurations with vorticity $(0,\pm1)$ in
each plaquette.  The parameter $y_a$ can be understood
as a core energy for a kink of a domain wall into one
layer from the other, and as has been discussed elsewhere \cite{herb2}
should be taken proportional to $1/h$ to match the energetics
of such kinks in the domain wall model $F_D$.

To identify the phases of the vortex system, we wish to find
the fixed points to which $H_{eff}$ flows under the RG.
We write the effective Hamiltonian as $H_{eff}=H_0+H_1$, with the
unperturbed Hamiltonian
\be
H_0=\int d^2 r \f{1}{2\al}|\na \phi_1+a\hat{x}|^2+\f{1}{2\al}
|\na\phi_2-a \hat{x}|^2+\f{1}{2h}(\f{\pa a}{\pa y})^2
+\f{1}{2}\rho a^2
\ee
with initial value $\rho(\ell=0)=2\pi y_2(\ell)$, and
\be
H_1=(-y_{\phi_1} \cos2\pi \phi_1+y_{\phi_1})+(-y_{\phi_2} \cos2\pi \phi_2 +y_{\phi_2})
+\sum_{n=2}^{n=\infty}\f{y_{2n}}{2n!}(-1)^n(2\pi a)^{2n}
\ee
with the initial values $y_{2n}(\ell=0)=y_a$.  (Here $e^{\ell}$ represents
the length scale to which the degrees of freedom have been integrated
out in the RG.)
We perform the RG perturbatively in $H_1$; since $y_a \sim 1/h$,
this amounts to looking at the RG flows in the region of a 
strong coupling fixed point.

To lowest order in $y_{\phi_1},~y_{\phi_2},$ and $y_a$, the RG flow 
equations can be shown to take the form \cite{herb2}
\bea
\f{dy_{\phi_1}}{dl}=(2-\pi K\sqrt{\f{1+K\rho}{K\rho}}) y_{\phi_1} \\
\f{dy_{\phi_2}}{dl}=(2-\pi K\sqrt{\f{1+K\rho}{K\rho}}) y_{\phi_2} \\
\f{dy_{2n}}{dl}=-(2n-2)y_{2n}-2\pi^2 L(\rho,\xi)y_{2n+2} \\
\f{d\rho}{dl}=-8\pi^4 L(\rho,\xi)y_4 ,\\
\label{flow}
\eea
where (for $\rho <<1$)
\be
L=\f{K\xi}{\pi\sqrt{K\rho(1+\xi\Lambda)}}
\ee
with $\Lambda=\pi/a$ and $\xi=\sqrt{K/h}$.
We note that since $\rho$ is small throughout the
flow range, the vortex fugacities $y_{\phi}$ and $y_{\theta}$ 
are strongly irrelevant, indicating that vortices are
always bound in both layers for the parameter regime
where our calculations apply, large vortex core energies
and small $1/h$ (i.e., {\it large} symmetry-breaking field.)
This may seem at first surprising, since we expect
that ``symmetric'' vortices (i.e., the combination
$M+N$ in vortex free energy, Eq. \ref{fv}) should
undergo a Kosterlitz-Thouless transition.  This is true,
but in our formulation it can be found only at order 
$y_{\phi_1}y_{\phi_2}$ because this involves creating
vortices in {\it both} layers at a given position.
In the domain wall representation, this corresponds
to an operator of the form 
$y_{\phi_1+\phi_2}\int d^2r \cos[\phi_1({\bf r})+\phi_2({\bf r})]$
becoming relevant, so that closed ``double'' domain walls -- closed
domain walls lying in both layers at the same time -- cannot
proliferate for the parameters corresponding to the unbound
symmetric vortex state.  At our order in perturbation
theory this has no effect, nor do we expect it to qualitatively:
since ``single'' domain walls (i.e., closed domain walls in
a single layer, generated by either $\phi_1$ or $\phi_2$) are
proliferated, the presence or absence of these higher order
domain walls should not affect the qualitative physics of
single vortex unbinding.

The ``mass'' parameter $\rho$ plays a crucial role in
determining the meaning of the fixed point Hamiltonian.
If $\rho \rightarrow 0$ then the field $a$ may effectively
remove arbitrarily large sections of the closed domain wall sections
generated by fluctuations in the fields $\phi_1$ and $\phi_2$.
This corresponds to an unbound phase of dislocations.
When $\rho>0$, it becomes prohibitively expensive energetically
to remove large domain wall segments, and the dislocations
remain bound.  Thus, the initial parameters that divide
trajectories for which $\rho(\ell \rightarrow \infty)>0$
from those in which $\rho(\ell \rightarrow \infty)=0$
represent a boundary separating
states in which dislocations are paired or
deconfined \c{herb1,herb2}.  The flow equations \ref{flow}
have been studied in detail in Refs. \onlinecite{herb1} and
\onlinecite{herb2},
and they demonstrate that a transition from $\rho=0$ to $\rho>0$
at the fixed point indeed occurs for a critical value of
$y_a$; i.e., a critical value of $h$.  Thus, for arbitrarily
small vortex fugacity, there is a dislocation unbinding transition
at a critical value of the symmetry-breaking field, $h_c$, with
the unbound state on the small $h$ side of this, and the bound
state on the large $h$ side.

Because of the duality between
vortices and dislocations, {\it this implies there must also
be a vortex unbinding transition.}  The duality suggests that
the unbound vortex phase should occur for small $h$ and small $E_c$,
as one might intuitively expect.
There are then three
possible phases for the system, an unbound vortex phase and two
bound phases, the latter being distinguished by whether
dislocations are in a bound or unbound state.  As has been
discussed in Ref. \onlinecite{herb2}, these two phases
may be described in terms of the vortex degrees of freedom
by whether the VAV pairs are linearly confined, or logarithmically
bound.  The linearly confined phase may be understood as the
natural situation for very dilute vortices.  Returning to the
original $XY$ model with a symmetry-breaking field,
a state with a single high separated VAV pair has its energy
minimized by creating a string of overturned angle with finite
width of order $\xi$ connecting the two topological defects \c{gm}.
(For separations smaller than this, the interaction energy is approximately
logarithmic, as in the absence of the symmetry-breaking field.)
This string arises because the symmetry-breaking field introduces
a large energy cost for states in which the phase deviates from $\theta=0$
over a large area.  Since a vortex requires a $2\pi$ angular rotation
for any path surrounding its core, one minimizes the loss in interaction
energy with the field by confining the rotation to a narrow region.
These strings may be understood as a degree of freedom dual to the
domain walls that arose in the dislocation representation.

The interpretation of the $\rho>0$ phase comes in part from noticing
that $y_{\phi}$ is irrelevant, so that domain walls are proliferated
by thermal fluctuations.  Because of the duality, we expect the same
to be true for the strings in the vortex representation of this state.
Physically, this means the entropy of the strings overwhelms their 
energy, so that strings of arbitrarily large size may be found in
a typical configuration.  However, the {\it logarithmic} attraction
between VAV pairs is not screened by the proliferating strings, so
they remain bound within the intermediate phase.  Vortex
unbinding only occurs at higher temperatures and/or lower $E_c$ and/or
lower $h$ when the entropy overcomes the logarithmic interaction.
A detailed discussion of how the RG analysis leads to this picture
in the single layer case may be found in Ref. \onlinecite{herb2}.

Finally, we note that the existence of all three phases has
been demonstrated in the $XY$ model with a magnetic field
using a simulation method that directly
probes the fluctuations of the vortices.

\section{Vortex Effects in Counterflow Resistance}

In this section, we
discuss how the different vortex phases could be distinguished
in a transport experiment.  Because the system will contain
``double'' vortices (pairs of vortices stacked across the layers)
which behave in a fashion similar to
vortices in a single layer system, deconfinement
of single vortices would be difficult to see in
a measurement the $I-V$ curve of the bilayer film
as a whole.  The problem of competing effects due to double
vortices can be overcome if we consider a {\it counterflow}
current, which creates forces on a single vortex but
not on a double vortex.
We thus consider a geometry
in which the current J is injected and removed  in
the $\hat{x}$-direction
from the two layers at the same edge (see Fig. 1). 
The current distribution
in such an experiment has been analyzed in the absence of
vortices and other fluctuations \c{cur}, where it was found that
the current within each layer $j_x$ and the interlayer (tunneling) current 
$j_z$ behave as $j_x \sim  sech(x/\lambda_J)$  and
$j_z \sim sech(x/\lambda_J) tanh(x/\lambda_J)$ with $\lambda_J = \sqrt{K/h} $
the Josephson length.
These currents are non-vanishing within a strip
of width $\sim \lambda_J$, so that vortices entering this region
will be subject to forces due to the current.  To simplify the
analysis below, we will model this by a uniform
tunneling current within $\lambda_J$ of the edge, and take the
currents to be zero deeper inside the sample.

To understand qualitatively the impact of the vortices,
consider one vortex in the relative phase field $\Phi=\theta_1-\theta_2$.
This configuration can be realized as $(M,N)=(1,0)$ or (0,-1), where $(M,N)$ is
the vortex number in each layer (see Eq. \ref{fv}).
For $(M,N)=(1,0)$, there is one vortex in the
upper layer, and none in the lower layer. 
Suppose the current is directed in the lower layer so that the
resulting motion of the vortex generates
a voltage $+V$ relative to that in the middle of the system,
where we choose the electric potential to be zero. For a single vortex of the
form $(M,N)=(0,-1)$, the vortex will move in the same direction
as in the previous case because both the direction of current in
the relevant layer and the vorticity have changed sign.  Since this
creates a vortex current of the {\it opposite} sign of that found
in the upper layer, the voltage generated in that layer will
have the same magnitude on average but opposite sign, $-V$. 
Thus, any vortex current induced by the counterflow current
produces an interlayer voltage at the edge.

To analyze this in detail, we
adopt the approach of Ambegaokar et al. described
in Ref. \onlinecite{ahns}, which
we hereafter refer to as AHNS.
AHNS demonstrated that the dynamics of vortices in response
to a driving current, and the voltage they induce, may
be described using a Fokker-Planck equation.
In this approach,
the separation $r$ of a VAV pair 
is described by the stochastic differential
equation
\be
\f{dr}{dt}=-\f{2D}{k_B T}\f{\pa U}{\pa r} +\eta(t) ,
\ee
where $U$ is the effective potential
for a VAV pair,
$D$ is the diffusion constant, and $ \eta(t)$ represents the thermal noise, with
the correlation function $<\eta(t)\eta(t')>=4D\de(t-t')$.
Since the vortices of interest
are located in a strip of width $\lambda_J$ near the system edge,
we will in fact solve a one dimensional problem. From the above stochastic
differential equation, one obtains a Fokker-Plank equation, describing
the relation between $\Gamma(y)$, the number density of VAV pairs 
near the edge with vertical separation
$y$, and the vortex current density $I$. 
One may start from
\be
\G(y,t+\De t)=\int dy' P(y,t+\De t|y',t)\G(y',t),
\label{G}
\ee
where $P(y,t+\De t|y',t)=<\de(y-y(t+\De t)>_{y't}$ is the probability for
a VAV pair having separation $y$ at time $t+\De t$, given that it had
separation $y'$ at time $t$. From Eq. (12) one has
\be
y(t+\De t)=y'-\f{2D}{k_B T}\f{\pa U}{\pa y'}\De t+\int_t^{t+\De t}\eta(t_1)dt_1
\ee
Expanding $P(y,t+\De t|y',t)$ to first order in $\De t$, we obtain
\be
P(y,t+\De t|y',t)=\Biggl[1-\f{2D}{k_B T}\f{\pa U}{\pa y'}\De t \f{\pa}{\pa y'}
+\f{1}{2}\int_t^{t
+\De t}dt_1 \int_t^{t+\De t}dt_2 <\eta(t_1)\eta(t_2)>\f{\pa^2}{\pa y'^2} \Biggr]\de(y-y')
\ee
Making use of the fact that $\int_t^{t+\De t} dt_1 \int_t^{t+\De t} dt_2
<\eta(t_1)\eta(t_2)>=4D\De t$ and integrating by parts in Eq. \ref{G}, one can obtain
\be
\f{\pa\G}{\pa t}=\f{\pa}{\pa y} (\f{2D}{k_B T}\G\f{\pa U}{\pa y}+
2D\f{\pa\G}{\pa y}),
\ee
which may be rewritten as 
\be
\f{\pa\Gamma(y)}{dt}=-\f{\pa}{\pa y}I.  
\ee
with
\be
I=-2Dexp(\f{-U}{k_B T})\f{\pa}{\pa y}[\Gamma(y)exp(\f{U}{k_B T})].
\label{fpI}
\ee
For a steady-state solution, $I$ is a constant and has the interpretation
of the number of separating pairs per unit time; i.e. the generation rate
for free vortices. The concentration of unbound
vortices $n_f$ is determined by the balance equation
\be
\f{dn_f}{dt}=I-\beta n_f^2.
\label{nf}
\ee
In this equation, the first term represents a generation rate
for unbound VAV pairs, while the second term
is
due to vortex recombination. In steady state ($\f{dn_f}{dt}=0$),
$n_f\sim \sqrt{I}$. 
Finally, we note that the interlayer voltage is generated
by a net vortex current, and so
is proportional to the density of free vortices.
Thus we have $\rho \sim n_f \sim \sqrt{I}$.
Our task is to compute $I$. From Eq. \ref{fpI}, we have
\be
I \propto \f{\Gamma(y_0)e^{U(y_0)}-\Gamma(L)e^{U(L)}}
{\int_{y_0}^L dy e^{U/K_BT}},
\ee
where $L$ is the system size and $y_0$ is a short distance cutoff.
At low temperature, to form a VAV with large separation, there is a large
energy barrier to overcome. Thus to leading order in $1/L$, we 
expect $\Gamma(L\rightarrow \infty)=0$.
So we have
\be
I \propto \f{1}{\int_{y_0}^L dy e^{U/K_BT}}.
\label{if}
\ee
Because there are three phases in this system, there are three
effective potentials and resulting resistances to consider.

\subsection{Linearly Confined Phase}

For the weakest fluctuations (lowest temperatures), we may
ignore any screening effects of the vortices on their mutual interactions.
As discussed above, this leads to a phase in which a VAV pair is
effectively connected by a string of overturned phase, generating an
interaction for highly separated VAV pair that grows
linearly with separation.  
An effective potential that describes this situation in the presence
of a current $j$ is
\bea
U=\va_0 ln(r/a)+\va_1r/a-\g jr  \\
=\va_0ln(r/a)-\De r  ,
\eea
where $\De=\g j-\va_1/a$,  $\va_1$ is the string free energy per unit length,
and $j$ is the current.  The $\gamma jr$ term is introduced to account
for the force on a vortex due to the current.  Because of this,
there is a critical current $j_c=\f{\va_1}{a\g}$.
There are two cases to consider:

{\it (i) $j>j_c$, $\De>0$.}  In this situation
there is a critical VAV separation $r_c=\va_0a/\De$, for which
the potential $U(r)$ increases with increasing r for $r<r_c$, and decreases
with increasing r for $r>r_c$.
VAV pairs
with separation shorter than $r_c$ tend to shrink, while VAV pairs
with separation larger than $r_c$ tend to increase the separation, eventually
becoming free vortices. For system size $L>>r_c$, a saddle point
approximation leads to
\bea
I\sim e^{-U(r_c)/k_BT} \\
\sim\De^{\va_0/k_BT} .
\eea
From Ohm's law, the interlayer voltage generated then takes the form
\be
V=\rho j \sim j(j-j_c)^{{\va_0}/{2k_BT}}.
\ee

{\it (ii) $j<j_c$, $\De<0$}.
In this regime, the potential $U(r)$ increases monotonically to infinity.
Thus for infinite system, the VAV pairs have to overcome an infinite high
barrier to become free vortices. We expect no free vortices
for an infinite system.  In finite size systems, we expect corrections
to this, which may be estimated as follows.
\bea
I\sim \f{1}{\f{k_B T}{-\De}(\f{r}{a})^{\va_0/k_B T}
e^{-r\De/k_B T }|_{x_0}^L+
\int_{x_0}^L e^{-r\De/k_B T }(\f{r}{a})^{(\va_0/k_B T-1)}dr}\\
\sim\f{1}{\f{k_B T}{-\De} (\f{L}{a})^{\va_0/k_BT}e^{-\De L/k_B T}  } \\
\sim \f{-\De}{k_BT}(\f{a}{L})^{\va_0/k_BT}e^{\De L/k_BT}
\eea
and
\be
V=j\rho\sim j\sqrt{\f{(j_c-j)}{k_B T}}(\f{a}{L})^{\va_0/2k_BT}e^{-\g(j_c-j)L/2k_BT}.
\ee
When system size
goes to infinity, the interlayer voltage disappears
throughout the region $j<j_c$, 
and {\it we have true dissipationless
superconductivity}. This is in contrast to single layer superconductors,
which generically have a power law $I-V$.
For current very small $\sim 0$, the resistivity
decreases with increasing of system according to exponential law,
i.e. $\rho \sim e^{-\va_1 L/2ak_BT}$.  We will demonstrate this
behavior in our simulations below.

\subsection{Logarithmically confined phase}

In this phase, the potential for a VAV pair has the form
\be
U(r)=\va_0 ln(r/a)-\g jr .
\ee
There is again a critical VAV separation $r_c=\f{\va_0}{\g j}$
above which they are free. 
For very large system size $L>>r_c$, we can apply a saddle point
approximation to the integral in Eq. \ref{if}
and we find
\be
V\sim \rho j \sim j^{1+\va_0/2k_BT} .
\ee
This is exactly the result of AHNS.
For fixed finite system size L, we consider the situation of very small
current $j \rightarrow 0$ to obtain an Ohmic dissipation,
\bea
I\sim \f{1}{\int_{y_{0}}^L (\f{r}{a})^{\va_0 /k_B T} }\\
=(\va_0/k_BT+1)(\f{a}{L})^{-(1+\va_0/k_BT)} .
\eea
Thus the system size dependence of the resistivity follows a power law 
with system size,
$\rho \sim L^{-(1/2+\va_0/2k_BT)}$.

\subsection{Free vortex phase}

For the free vortex phase, in the limit of infinite system size,
the density of free vortices is a constant, and we get a linear 
$I-V$; i.e., Ohm's law. For finite system size, some bound
VAV pairs with separation of order $L$ also contribute to the
voltage. Thus with increasing $L$, the resistivity decreases (due 
to the decreasing number of effectively
unbound VAV pairs) and saturates to a non-zero value as $L \rightarrow \infty$.

\section{numerical simulation}
To test the results discussed above, we studied the dissipation due to
vortices in Langevin dynamics simulations. The simulation
models a Josephson coupled
bilayer superconductor, in which we directly simulate
{\it only the relative phase degree of freedom}.  This is an important
simplification because it cuts in half the computer time needed to collect
data for a given set of parameters, which is often considerable.
The dynamics of the angles in this system are simulated by integrating
the classical equations of motion, so that our system may be
understood as a variation of the resistively shunted Josephson
junction model \c{sk}. As described below,
we introduce
explicit, ideal (dissipationless) leads, which are weakly coupled to each of the
two edges
of the system at
$x=\pm L_x/2$.  In the $y$-direction we adopt periodic boundary conditions.
In our Langevin dynamics simulation, the equations of
motion are taken to be
\be
\G\f{d^2 \Phi(r)}{dt^2}=\f{\de H_{XY}}{\de\Phi(r)}+\zeta(r)-
\eta\f{d\Phi(r)}{dt}.
\ee
Here  $\Phi(r)$ include the phase difference for the two layers and the
two leads. Following Ref. \onlinecite {sk}, we apply a 
``busbar geometry'' in which there is a {\it single} phase
variable in each of the leads.  This eliminates the possibility of
vortices in the leads themselves, and allows us to focus on
dissipation due only to the system.  In practice, this
might be accomplished by using superconducting films for
the leads that are much thicker than those of the system,
so that they will have a much higher superfluid stiffness \c{girv_com}
than inside the system itself.  
An important aspect of the simulation is that the coupling 
between the system and the leads must be {\it weak} to
avoid disturbing the behavior of vortices that approach the
edge.  
For the results reported here, we took $K_L=K_R=0.05$ (see below)
whereas the coupling within the system was taken as $K=1$, setting
our unit of energy.
The effective moment of inertial for the spins were also taken to
be 1. The viscosity
$\eta$ was taken to be 0.143 in our simulation.
$\zeta(r)$ is a random torque satisfying $<\zeta(r,t)\zeta(r',t')>=2\eta
T\de_{r,r'}\de(t-t')$ with $T$ being the temperature of the system (chosen
to be 1.2 for the results discussed below.)
For a system of size $N_x,N_y$,
our Hamiltonian $H_{XY}$
takes the form
 \bea
H_{XY}=-K\sum_{<r,r'>}cos[\Phi(r)-\Phi(r')]-h\sum_r cos[\Phi({\bf r})] \\
- K_L \sum_{j=1}^{N_y} cos(\Phi_L-\Phi(a\hat{x}+aj\hat{y}))   -K_R \sum_{j=1}^{N_y}
cos(\Phi_R-\Phi(N_xa\hat{x}+aj\hat{y}))
 \eea
 with $K=1$.
A typical run consists of $3.9\times 10^8$
time steps. Each time step is 0.08 (in the unit of $\sqrt{\G/K}$). 
We also eliminate the initial
$1.5\times 10^7$ steps in these runs for equilibration. We repeated runs for each set of
parameters with several different seeds, allowing us to estimate the
statistical error.

In principle, we can find the resistances and I-V curves by injecting
counterflow current, which in this model would act as a constant force
on the lead phase variables $\Phi_L$ and $\Phi_R$.  The induced interlayer voltage
is then just given by the Josephson relation, $V_{L,R} \propto
d\Phi_{L,R}/dt$.  However, for low currents such simulations become
very challenging because there are long time scales involved.
To simplify our calculation, instead of calculating
the $I-V$ curves directly, we calculate the resistance for small current
($j\rightarrow 0$) via the Einstein relation. The response of a
lead variable $\Phi_{L,R}$ to a DC driving force (current)
is proportional to its diffusion constant, which can be calculated by
the correlation function
\be
D=\lim_{T \rightarrow \infty} \f{1}{T}<(\Phi'(t+T)-\Phi'(t))^2>_t ,
\ee
where, for example, $\Phi'=\Phi_L-\Phi_0$, and
$\Phi_0$ is a spin angle at some point deep inside the system. Here we choose it to
be the middle point of the sample. $<...>_t$ refers to an average over different
starting times, which is essentially an average over initial conditions.
Because in our model current acts as a driving force and voltage
is an average velocity of the lead phases in response,
the diffusion constant for the lead variables are proportional to
the {\it resistance} of the system.

Our simulation results are summarized in Fig. 2. We show the system
size ($N_y$) dependence of the diffusion constant (i.e. resistance) for $h$=0.0286,
0.143,1.43. Our system size is $N_x=29$ (fixed) and $N_y$=40,50,60,75,85.
It is clear that there are three different behaviors. The curve with solid dots
is for $h=0.0287$, which shows a straight line with slope very close to -1. Here the
dashed line is a reference line with slope exactly -1. The resistance $R\sim N_y^{-1}$
comes purely from the geometry of the sample and
implies a constant {\it resistivity} (i.e., $R=\rho N_x/N_y$). 
Thus this result is consistent with the system being in the free 
vortex phase.
The curve with solid squares for $h=0.143$ is a straight line with slope $>1$
on the log-log plot of Fig. 2, indicating 
that the system size dependence of the resistivity has a power law behavior.
This is consistent with our expectations for the
logarithmically confined vortex phase. Finally, the
curve with solid triangle for $h=1.43$ shows 
a marked downward curvature on the log-log plot.
In the insert, this curve is shown in a log-linear plot, demonstrating a nice
exponential law for the system size dependence.  This is consistent
with our expectations for the 
linearly confined phase. Thus our numerical results show that for a fixed
temperature $T>T_{KT}$, with increasing $h$ (interlayer tunneling), 
the resistance has three
different possible qualitative behaviors. 

\section{Conclusion}
By a combination of analytical analysis and numerical simulation, we have shown that
a bilayer thin film superconductor supports 
three vortex phases for the antisymmetric combination
of the layer phase variables: a free vortex phase, a logarithmically confined
phase, and a linear confined phase. 
We argued from a Fokker-Planck analysis that the
corresponding interlayer
$I-V$ curves should show measurably different behaviors: 
metallic conductivity for the unbound phase, a power law
$I-V$ in the logarithmically bound phase, and true 
dissipationless superconductivity (for an infinite system)
in the linearly confined phase.  We demonstrated that this behavior
is consistent with what is found in numerical simulations.

{\bf Acknowledgments} The authors would like to acknowledge
discussions with Steve Girvin in the early stages of this work.
This research was supported by NSF Grant Nos. DMR-0454699 and DMR-0511777.

\newpage
\vfill\eject
\begin{figure}[b]
\includegraphics[width=10cm]{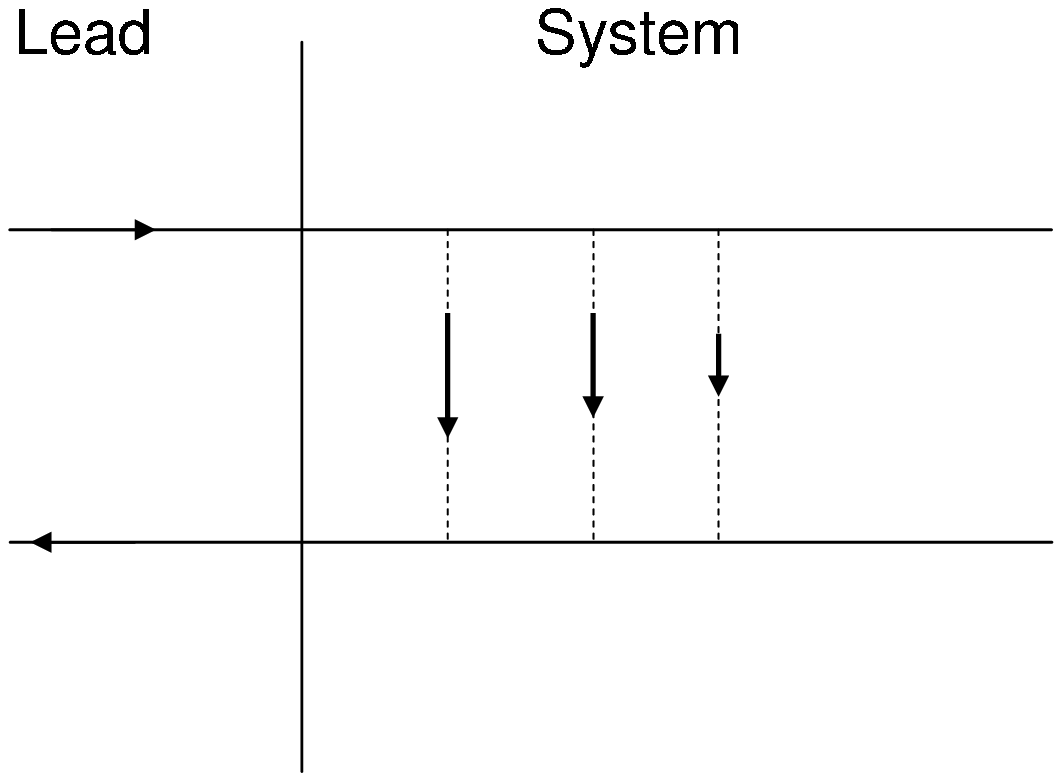}
\caption{Bilayer system with current being injected
and removed from the two layers at the same edge by an ideal lead.
}
\end{figure}
~~~~~~~~~~~
\vfill\eject

\newpage
\vfill\eject

~~~~~~~~

\begin{figure}[b]
\includegraphics[width=15cm]{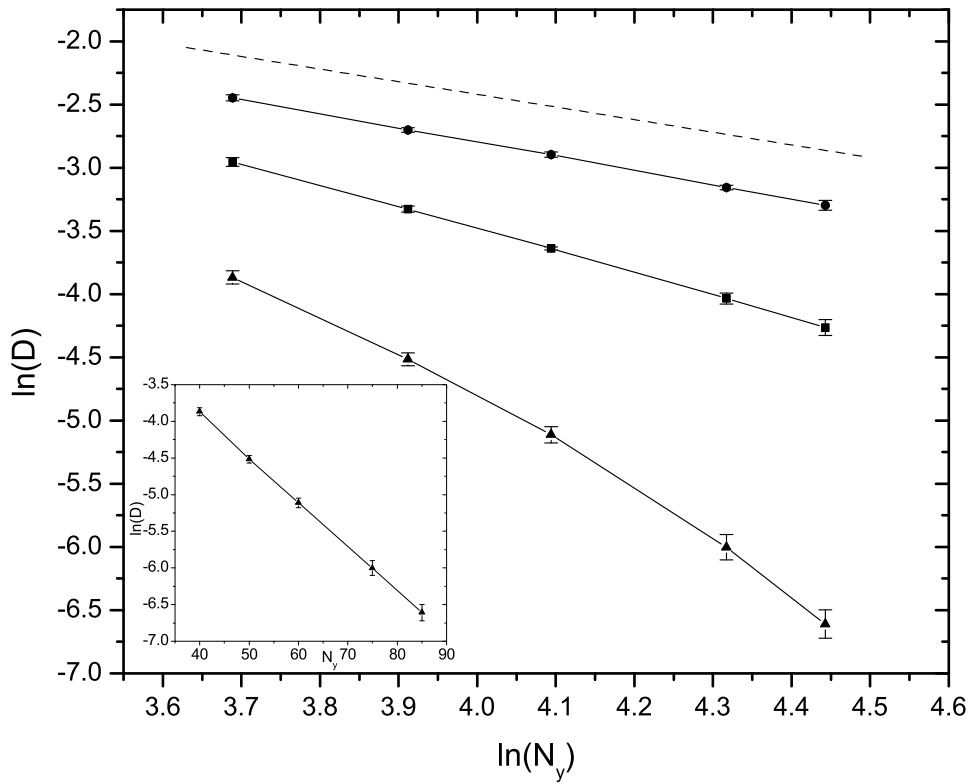}
\caption{The diffusion constant vs. system size $N_y$ in a log-log plot
with error bars. $N_x=29$ is fixed. The curve with solid dots is for
$h=0.0286$, corresponding to the free vortex phase, showing $D(\sim R)\sim
N_y^{-1}$. The dashed line is the reference line with slope -1.  The curve
with solid squares is for $h=0.143$, corresponding to logarithmically confined
phase, showing $D(\sim R) \sim N_y^{-a}, a=1.74>1$,
a power law $N_y$ dependence.
The curve with solid triangles is for $h=1.43$, corresponding to linear confined
phase. The insert shows the data for $h=1.43$ in a log-linear plot,
clearly demonstrating the exponential system size
dependence.
}
\end{figure}

~~~~~~~~~

\vfill\eject

\end{document}